\documentclass[superscriptaddress,floatfix,amsmath,amssymb]{revtex4}
\usepackage{graphicx}
\usepackage{amssymb}
\usepackage{epstopdf}
\usepackage{color}
\usepackage{bm}
\usepackage{graphicx}

\begin{document}
\title{Statistical Properties of  Many Particle Eigenfunctions }
\author{Eric J. Heller}
\email{heller@physics.harvard.edu }
\affiliation{Department of Physics, Harvard University, Cambridge, MA 02138}
\affiliation{Department of Chemistry and Chemical Biology, Harvard University, Cambridge, MA 02138}
\author{Brian R. Landry}
\affiliation{Department of Chemistry and Chemical Biology, Harvard University, Cambridge, MA 02138}

\date{\today}

\begin{abstract}
Wavefunction correlations and density matrices for few or many particles are derived from the properties of semiclassical energy Green functions. Universal features of fixed energy (microcanonical) random wavefunction correlation functions appear which reflect the emergence of the canonical ensemble as $N\to\infty$.  This arises through a  little known asymptotic limit of Bessel functions. Constraints  due to symmetries, boundaries, and collisions between particles can be included. 
 \end{abstract}
 
\maketitle
\section{Introduction}
The standard tools of quantum chaos investigations include random matrix theory and periodic orbit theory (Gutzwiller trace formula), the Van Vleck-Morette-Gutzwiller propagator,  and many techniques and phenomena derived from these approaches. Standing somewhat to the side as an inspired insight is Berry's  conjecture, which loosely stated is the idea that as $\hbar \to 0$ eigenstates will be indistinguishable from superpositions of infinitely many (local) plane waves with random amplitude, direction, and phase, but with fixed wavelength appropriate to the local kinetic energy.  In two dimensions, these assumptions result in strictly Gaussian statistics of the eigenfunctions and the  autocorrelation function 
$ 
\langle \psi^*(\vec x)\psi(\vec x +  \vec R )\rangle = J_0(k a) 
$
where $k$ is the local wavenumber and $\vert  \vec R \vert = a$.

 The Berry random  plane wave (RPW)\cite{berry1} hypothesis is free of any specific dynamical information, except  fixed total energy, which defines the ``ensemble'' (i.e. microcanonical). The perspective  developed here suggests that by extending the RPW hypothesis we can conveniently accommodate many other constraints, incorporating  information about real systems.  In fact this program has already begun, with Berry's inclusion of the presence of nearby hard walls\cite{berrywall}, and Bies and Heller's soft boundary results\cite{biessoft}, and multiple hard walls\cite{cone}.  Related work by  Urbina and Richter\cite{richter1} and one of us \cite{stat1} may also be viewed in this light.
 
 The   idea of  random waves subject to constraints is not confined to one particle in two dimensions.   Indeed Berry gave the $N$ - dimensional formula for free particles  in his 1977 paper\cite{berry77}.  Since the underlying idea in the RPW hypothesis is uniform randomness within a quantum context, i.e. the underpinning of quantum statistical mechanics, we  must encounter some familiar territory  as   the RPW hypothesis is extended to the large $N$ limit.   In 1994, Srednicki had suggested that the Berry random wave hypothesis was indeed a foundation for quantum statistical mechanics\cite{srednicki}, and showed that the appropriate canonical ensemble was reached for large $N$, depending on particle statistics. The present paper shows  more specifically what happens as the number of particles increases, through a nonstandard and apparently unpublished asymptotic form for Bessel functions (we have not been able to find it in the literature, although it ``ought'' to be there), which encodes the equivalence of the canonical and microcanonical ensembles of statistical mechanics. In making the connections to quantum statistical mechanics one also needs procedures for incorporating constraints, which are an essential aspect of the theory.  Thus our procedures for generalizing the RPW to include constraints, mentioned above,  is an essential new feature, since the constrained eigenstates are no longer random in Berry's (and Srednicki's) original sense.   



Given a  continuum at energy $E$, such as in an enclosure with walls very far away, we can perform the average over all random waves as a trace, i.e.
\begin{equation} 
\langle \psi^*(\vec x)\psi(\vec x'  )\rangle ={\rm Tr} \left [\delta(E-H) \vert \vec x \rangle\langle\vec x'   \vert\  \right ],
\end{equation} 
which immediately yields Berry's result, apart from normalization which we choose differently here.   However
a trace over a basis is independent of any unitary transformation on that basis, so it does not matter whether we use a trace over a complete set of random waves or simple local plane waves; both give $J_0(k a )$  for the case of one  free particle in two dimensions. In this way the imaginary part of the retarded Green's function  $-\frac{1}{\pi}{\rm Im}\left [ G^+(E) \right ]=\delta (E-H)$ becomes central, formally convenient, and equivalent to Berry's RPW hypothesis.

\section{Preliminaries}

 We begin by reviewing well known formalism to establish context and notation. The Green function completely characterizes a quantum system, whether it is interacting or not,  or has few or many degrees of freedom.  The retarded Green function $G^+$, i.e. 
\begin{equation} 
G^+ = {\cal P} \frac{1}{E-H} - i \pi\delta (E-H),
\end{equation} 
where  ${\cal P} $ stands for the principal value of the integral, is the basis for wavefunction statistics and density matrix information, through the follow relations, with a convenient choice of normalization:
\begin{eqnarray} 
<\psi ( {\bf x}) \psi^*( {\bf x'})> & = & -\frac{1}{\pi} {\rm Im}\langle {\bf x}\vert G^+\vert {\bf x'}\rangle/\rho(E)\\&=& \langle {\bf x}\vert\delta(E-H)\vert {\bf x'}\rangle  /\rho(E)
\label{greencor}
\end{eqnarray}  
where 
\begin{equation} 
\rho(E) = {\rm Tr}[\delta(E-H) ]
\end{equation} 
and where   $<\cdots>$ stands for the average over the degeneracies. We take these degeneracies to be of dimension up to $ND-1$, where $N$ is the number of particles and $D$ the spatial dimension each particle lives in. (We use boldface notation, e.g. ${{\bf x}}$ for the $N*D$ degrees of freedom.)   
If true degeneracies do not exist in a particular system, we can artificially open the system up to a continuum. For example, a two dimensional closed billiard does not have a degeneracy, but it acquires one if we open a hole in it and let it communicate with the outside unbounded 2D space.  Of course this changes the billiard properties, and the size of the hole might be problematic, but in fact we shall never really have to open a system up in this way.  The quantity  $\delta(E-H)$ then implies the average over all scattering wavefunctions at fixed energy $E$. 

There are other  interpretations which can be put on the average correlation $<\psi ( {\bf x}) \psi^*( {\bf x'})>$; for example we can imagine a large number of potentals which differ in some far away place, and in a way so as to all have an eigenvalue at a particular energy.  Then, the average has the interpretation of the average over this ``disorder'' ensemble.  A slightly different procedure is advocated by Richter {\it et. al.}, wherein an energy average is taken\cite{richter1}. Another interpretation can be applied to individual eigenstates in a closed system, assuming they are at least locally uniform in their properties, by taking the average over different points of origin ${\bf x}$. This is particularly appropriate when the analogous classical system is chaotic, as mentioned above \cite{berry1}.   We will  be evaluating the Green functions semiclassically in what follows, restricting the time over which the contributing trajectories propagate. 

The wavefunction correlation is equal to the coordinate space matrix element    of the constant energy density matrix:
\begin{equation} 
<\psi ( {\bf x}) \psi^*( {\bf x'})> = \langle {\bf x}\vert\delta(E-H)\vert {\bf x'}\rangle /\rho(E) 
  =\rho ({\bf x},{\bf x'},E)
\label{density}
\end{equation} 
Reduced density matrices  can also be derived from wavefunction correlations ; e.g.
\begin{equation} 
\tilde \rho (\vec x_1,\vec x_1',E) = \int d\vec x_2 d\vec x_3\cdots d\vec x_N\  \rho (\vec x_1, \vec x_2,\cdots;\vec x_1^\prime,\vec x_2,\cdots;E),
\end{equation} 
the one particle reduced density matrix.

We can approach the correlations via Fourier transform from the time domain, since
\begin{equation} 
\delta(E-H) =\frac{1}{2 \pi \hbar} \int\limits_{-\infty}^{\infty}e^{iEt/\hbar}e^{-iHt/\hbar}\ dt.
\end{equation} 
Thus the statistics, density matrices and  correlations are derivable without further averaging by knowing the time propagator.  

In the following, we define the Green function propagator $G( {\bf x}, {\bf x^{\prime}},t)$ and the retarded Green function propagator $G^{+}( {\bf x}, {\bf x^{\prime}},t)$ as
\begin{eqnarray} 
G( {\bf x}, {\bf x^{\prime}},t)  &=&\langle {\bf x}\vert e^{-i Ht/\hbar}\vert {\bf x^{\prime}}\rangle \nonumber \\
G^{+}( {\bf x}, {\bf x^{\prime}},t) &=& {-i\over \hbar}\Theta(t)
\langle {\bf x}\vert e^{-i Ht/\hbar}\vert {\bf x^{\prime}}\rangle 
\end{eqnarray} 
where $\Theta(t)$ is the Heavyside step function
$\Theta(t)= 0$, $t < 0$, $\Theta(t)= 1$, $t > 0$.
It is very rewarding to expand the propagator in semiclassical terms, involving short time (zero length) and longer trajectories.  
We take $G_{direct}({\bf x},{\bf x}+ {\bf r},t) = \langle {\bf x}\vert \exp[-iHt/\hbar]\vert {\bf x}+ {\bf r}\rangle$,  the very short time semiclassical propagator, which for $N$ particles each in D dimensions reads
\begin{equation} 
\label{green}
G_{direct}({\bf x},{\bf x}+ {\bf r},t)  
\approx \left ( \frac{m}{2 \pi i \hbar t}\right )^{ND/2} e^{i m r^2/2\hbar t - i V({\bf x} +\frac{ {\bf r}}{2})t/\hbar}
\end{equation} 
where $r^2 = \vert {\bf r} \vert^2$.

 It is not difficult to cast the Fourier transform of this short time version to fit the definition of a Hankel function, i.e.
\begin{equation} 
 G_{cl}^+({\bf x},{\bf x}+ {\bf r},E)  =  \frac{-i}{\hbar}\int\limits_0^\infty\left ( \frac{m}{2 \pi i \hbar t}\right )^{ND/2} e^{i m r^2/2\hbar t - i V({\bf x} +\frac{ {\bf r}}{2})t/\hbar} e^{i E t/\hbar}\ dt = 
 -\frac{i m}{2 \hbar^2} \left ( \frac{k^2 }{2 \pi k r}\right )^d H_d^{(1)}(k r)
 \label{bes}
 \end{equation} 
 where $d = ND/2-1$, 
$ k = k({\bf x} + {\bf r}/2, E)$ and $H_d^{(1)}(k r)  = J_d(k r) + i N_d(k r)$ is the Hankel   function of order $d$, and $J_d$ is the regular Bessel function of order $d$. The wavevector $k$ varies with the local potential, i.e. 
$
 \hbar^2 k({\bf x},E)^2/2m = E-V({\bf x}).
$
Here, using only the extreme short time version of the propagator, we must suppose $ {\bf r}$ is not large compared to significant changes in the potential, but this restriction can be removed by using the full semiclassical propagator rather than the short time version.
 For the case of one particle in two dimensions, $d=0$, and we recover Berry's original result for one particle in 2D,  $\langle \psi^*(\vec x)\psi(\vec x +  \vec r )\rangle \propto  J_0(k r)$. 
 
 According to the short time approximation, for any $N$,
  \begin{equation} 
 <\psi ( {\bf x}) \psi^*( {\bf x}+{\bf r})> \approx   -\frac{1}{\pi} \frac{{\rm Im} \left [  G_{cl}^+({\bf x},{\bf x} +{\bf r}, E) \right ] }{\rho(E)}= \frac{1}{\rho(E)} \frac{m}{2\pi \hbar^2}\left (\frac{k^2}{2\pi k r}\right )^d\ J_d(k r)
\label{main}
   \end{equation}
 where   $k=k({\bf x},E)$.  This result includes interparticle correlations through the potential $V({\bf x})$ and the spatial dependence of $k = k({\bf x},E)$; the diagonal $r=0$  limit (following section) is equivalent to classical statistical mechanics. The implications of this for the nondiagonal short time Green's function are intriguing. The way $r$ is defined, it does not matter whether one particle is off diagonal (${\bf x}_i \ne {\bf x}_{i'}$)  or several or all of them.  For  given $r$, the Green's function will be the same, apart from changes in the potential $V({\bf x} + {\bf r}/2)$.

      It is interesting that although the short time Green function is manifestly semiclassical, the energy form, e.g. Eq.~\ref{main} is obtained by exact Fourier transform of the semiclassical propagator, rather than by stationary phase. 
         
 \section{Diagonal limit}
 \label{dl}
 
The diagonal ($ r \to 0$) 
N body Green function  is obtained 
using the asymptotic form
   \begin{equation} 
   \lim_{r\to 0} J_d(k r) =\ \frac{1}{\Gamma(d+1)} \ \left (\frac{ k r}{2 }\right )^d\approx  \frac{1}{\sqrt{ 2\pi d}}\left (\frac{e k r}{2 d}\right )^d
   \end{equation} 
   we obtain
    \begin{equation} 
    -\frac{1}{\pi}{\rm Im} \left [  G_{cl}^+({\bf x},{\bf x}, E) \right ]\approx \frac{m}{2\pi \hbar^2}\frac{1}{\Gamma(d+1)} \left ( \frac { k^2}{4 \pi }\right )^d\approx  \frac{m}{2\pi \hbar^2}\frac{1}{\sqrt{ 2\pi d}} \left ( \frac {e k^2}{4 \pi d}\right )^d
   \end{equation} 
   where the second form uses Stirling's approximation, $n! \sim n^n e^{-n}\sqrt{2\pi n}$, 
   and is appropriate below when we consider large $N$.
   We note that this behaves as $k^{2d} \sim (E-V(\vec x))^d$. This factor is familiar from the computation of the classical density of states.
   Tracing over all $\vec x$ results in 
\begin{eqnarray} 
\int d{\bf x}  \frac{m}{2\pi \hbar^2}\frac{1}{\Gamma(d+1)} \left ( \frac { k^2}{4 \pi }\right )^d &=& \int \frac{d {\bf x} d{\bf p}}{h^{ND}}   \delta (E-H_{cl}( {\bf p},{\bf x}))
 = \rho_{cl}(E)
\end{eqnarray}  
   i.e. the classical density of states.  The association of the short time propagator with the classical Hamiltonian and classical density of states is well known.
   The Berry RPW hypothesis, the short time propagator, and the classical or Weyl (sometimes called Thomas-Fermi) term in the quantum density of states are  all closely related.

The quantum spacial integral is over all coordinates, so how does the classical partition function emerge if the classical integral is only over classically allowed coordinates?  For forbidden positions,
 $k$ is imaginary and can be written as say $i \kappa$.  An identity for Hankel functions can then be used
 ($i^{n+1} H_n^{(1)} (i x) = \frac{2}{\pi} K_n(x)$) to show that the green function is real so that the imaginary part is zero, explaining why the integral is only over classically allowed positions. 
 
As long as $ {\bf r} =0$ (i.e. diagonal Green's function) the results obtained within the short time propagator approximation for any quantity in the presence of a potential ({\em including} interparticle potentials such as atom-atom interactions) will be purely classical.  Since we will be discussing the equivalence of the results from the different ensembles for $ {\bf r} \neq 0$, it is useful to recall how the classical coordinate space densities in the different ensembles can be shown to coincide since this corresponds to the $ {\bf r} =0$ case.

The normalized phase space density in the microcanonical ensemble and the phase space density in the canonical ensemble are given by 
\begin{equation}
\rho_{cl} ({\bf p},{\bf x} , E) =  \frac{1}{\rho_{cl}(E)} \ \delta (E-H_{cl}( {\bf p},{\bf x}))
\end{equation}
and
\begin{equation}
\rho_{cl} ({\bf p} , {\bf x}, \beta) = \frac{1}{Q_{cl} (\beta) } \ e^{-\beta H_{cl}( {\bf p},{\bf x})}
\end{equation}
respectively.  The density of states and partition function are of course the normalization factors so that
\begin{eqnarray}
\rho_{cl}(E) &=& \int d {\bf x} d {\bf p} \ \delta (E-H_{cl}( {\bf p},{\bf x})) \\
Q_{cl} (\beta) &=& \int d {\bf x} d {\bf p}\ e^{-\beta H_{cl}( {\bf p},{\bf x})}
 \end{eqnarray}
Integrating each phase space density over momentum space allows us to compare the coordinate space densities:
\begin{eqnarray}
\rho_{cl} ({\bf x}, E) &=&  \frac{p^{2d}}{\int d {\bf x} \ p^{2d}} \\
\rho_{cl} ({\bf x} , \beta) &=& \frac{e^{-\beta V({\bf x})}}{\int d {\bf x} \ e^{-\beta V({\bf x})}}
\end{eqnarray}
with $p = \sqrt{2m(E -V({\bf x}))}$.

Using the relationship between $E$ and $\beta$, $E - \left< V \right> = \frac{ND}{2 \beta}$, 
where $\left< V \right> $ is the ensemble average of the potential in one of the statistical ensembles, the coordinate space density becomes
\begin{eqnarray}
p^{2d} &=& (2m(d+1)/\beta)^d \left(1 + \frac{\left(\left<V\right> -  V( {\bf x})\right)\beta}{d+1} \right)^d
\end{eqnarray}
 In the limit $N \rightarrow \infty$ ( $d \rightarrow \infty$)  this is
 \begin{eqnarray}
p^{2d} &=& (2m(d+1)/\beta)^d e^{ \left(\left<V\right> -  V( {\bf x})\right)\beta} \\
\frac{ p^{2d} }{ \int d{\bf x} \ p^{2d} } &=& \frac{ e^{ - V\left({ \bf x}  \right) \beta} }{ \int d{\bf x} \ e^{ - V( {\bf x}) \beta} }
 \end{eqnarray}
This is one of the standard ways of establishing a connection between the ensembles\cite{jancel}.
   
Since the diagonal Green's function gives classical results we can use it to study classical properties.  For example, we can inquire about the average two particle spacing distribution $\rho_{E}(r_{12}) $ or the probability density for a single particle $P_E(\vec x_1)$ starting with the short time semiclassical Green's function and the results will coincide with classical microcanonical statistical mechanics.  This statement holds for all $N$. Similarly, in the large $N$ limit the canonical ensemble results for these quantities must emerge. This point becomes more interesting for the non-diagonal case, considered next.

 \section{Link to  the canonical ensemble}

 \subsection{Bessel functions become Gaussians}
 
\begin{figure}
\centerline {
\includegraphics[width=6in]{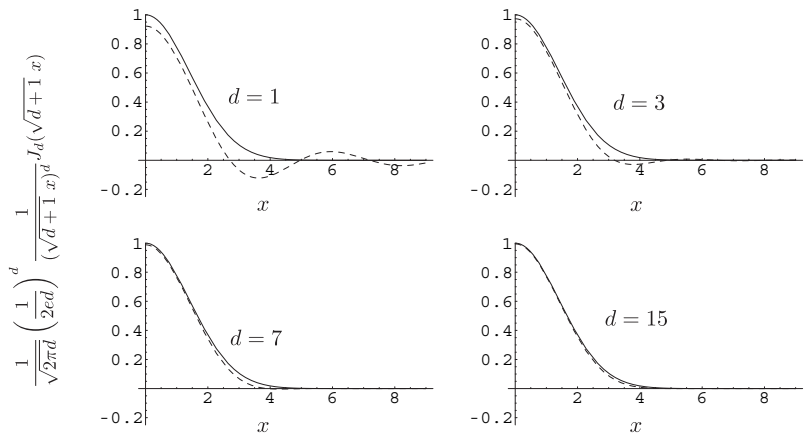}
}
\caption{As N increases, the combination $\frac{1}{x^d} J_d(x)$, where $d=ND/2-1$, approaches a Gaussian.  This is the key link between the quantum microcanonical and canonical ensembles.}
\label{bestoGauss}
\end{figure}

As yet we have found nothing too surprising or useful beyond standard classical statistical mechanics. This changes when we consider the large $N$ limit for the non-diagonal Green's function, $ {\bf r}\ne 0$. Taking the large $N$ limit of Eq.~\ref{main}, we are confronted with a new question about Bessel functions.  The large $d$ limit of $J_d(x)$ is indeed well known, but this is not yet sufficient for our purposes. It reads
   \begin{equation} 
   \lim_{d\to \infty} \frac{J_d(kr)}{(kr)^d} =\ \frac{1}{2^d\ \Gamma(d+1)} \ \approx  \frac{1}{\sqrt{ 2\pi d}}\left (\frac{e }{2 d}\right )^d
   \label{incon}
   \end{equation} 
  This is the standard formula given in the usual references. Eq.~\ref{incon} should be the first term in a power seres for  $J_d(kr)$ in $kr$. Another standard result is the power series expansion, valid for all $d$ and $kr$:
\begin{equation} 
J_d(kr) = \sum\limits_{m=0}^\infty\frac{(-1)^m}{m! \Gamma(m+d+1)}\left (\frac{kr}{2}\right )^{2m+d}
\label{asymp}
\end{equation}
 We actually require a different asymptotic result. What make our demands   unusual is that, assuming we want  the energy to increase in proportion  to the number of particles (appropriate to many applications of the 
large $N$ limit), then  $k\sim \sqrt{E} \sim \sqrt{N}\sim \sqrt{d}$; this means that for fixed $r$ the combination $(kr)$ is increasing as $\sqrt{d}$ as $d\to\infty$. If the argument of the Bessel function increases without bound along with it's order, some new considerations come into play. 
 We  find the desired form using Eq.~\ref{asymp}, after summing a series recognized as that of a Gaussian Taylor expansion,
\begin{equation} 
\lim_{d\to\infty}\frac{1}{(kr)^d} J_d(kr)= \frac{1}{2^d\ d!} \sum\limits_{m=0}^\infty \frac{1}{m!}\left (\frac{-k^2r^2}{4(d+1)}\right )^m = \frac{1}{2^d\ d!} e^{-k^2r^2/(4(d+1))},
\label{asympnew}
\end{equation} 
where again $ \hbar^2 k^2/2m = E-V({\bf x}).$
Note that as $d\to\infty$, the argument of the Gaussian holds fixed because of the factor of $d+1$ in the denominator of that argument.
 Figure~\ref{bestoGauss} illustrates the convergence to the Gaussian as $N$ increases. 
The asymptotic limit in Equation \ref{asympnew} is not in the usual references, although   related results have been given for N-bead polymer random chain end-to-end distributions\cite{kleinert}. The connection  between the path integral for the propagator and polymer chains is well known\cite{cw}.

  
It is interesting that a Gaussian emerges from Bessel functions in the large $N$ limit. We can put Eq.~\ref{asympnew} together with Eq.~\ref{main} and Eq.~\ref{greencor}, and express the result, as ${N\to\infty}$,
\begin{equation} 
<\psi ( {\bf x}) \psi^*( {\bf x}+{\bf r})> \ = \ \rho ({\bf x},{\bf x'},E)  \  \to \ \frac{1}{\rho(E)} \frac{m}{2\pi \hbar^2 d!}\left(\frac{k^2}{4 \pi}\right )^d\ e^{-k^2r^2/4(d+1)}.
\label{asymp1}
\end{equation} 
For noninteracting particles moving in zero potential  but confined to volume $V$ the short time approximation becomes exact and $k$ is constant.  For this system the wavefunction correlation becomes
\begin{equation} 
<\psi ( {\bf x}) \psi^*( {\bf x}+{\bf r})> \  = \ \rho ({\bf x},{\bf x'},E)  \  \to \  \frac{1}{V^N} \ e^{-k^2r^2/4(d+1)}.
\end{equation} 
Something familiar is emerging, here derived in the unfamiliar context of fixed energy (microcanonical ensemble).   
For comparison we recall the standard result for the ideal gas at temperature $T$\cite{pathria}:
\begin{equation} 
\frac{ \langle {\bf x} \vert e^{-\beta H}\vert  {\bf x}+ {\bf r}\rangle }{ {\rm Tr}[ e^{-\beta H}]} =  \rho ({\bf x},{\bf x'},\beta) = \frac{1}{V^N} \ e^{-\pi r^2/\lambda^2}
\label{path}
\end{equation} 
where $\lambda = h/\sqrt{2 \pi m \kappa T}$ is the thermal wavelength.
Indeed for the free particle case, $k$ is fixed by $E$ and  $\langle K \rangle = D/2N\kappa T = \hbar^2 k^2/2 m$, where $K$ is the kinetic energy and $\kappa$ is Boltzmann's   constant,
 \begin{equation} 
e^{-k^2r^2/4(d+1)} =  e^{-\pi r^2/\lambda^2}.
\label{ensembleequiv}
\end{equation} 
 
 
The canonical ensemble result  for the propagator has   ``dropped out'' of the asymptotic large $N$ limit  of a microcanonical Green function, at least for noninteracting particles, and   an unusual asymptotic form for the Bessel function has  emerged as  the link.  With some caveats, the statement 
\begin{equation} 
\delta (E-H) \sim e^{-\beta H}
\label{ensembles}
\end{equation} 
has meaning in the large $N$ limit, where it is understood $E$ grows as $N$, and a temperature extracted.  At a qualitative level, Eq.~\ref{ensembles} merely expresses the known equivalence of the ensembles.   In the case of an interaction potential, the relation between $E$ and temperature is of course problematical.

\subsection{Interacting Particles - Short Time Limit}
\label{stl}
We can say more about interacting particles using only  the short time propagator introduced above.  Longer time events will be discussed in Sec.~\ref{pot}.
The short-time approximation to the correlation function for large $N$, which is equal to the matrix elements of the density operator in coordinate space using our normalization, (Eq.~\ref{asymp1})  is given by
\begin{equation}
\rho_{cl} ( {\bf x},  {\bf x'}, E) = \frac{1}{\rho(E)} \frac{m}{2\pi \hbar^2 d!}\left(\frac{k^2}{4 \pi}\right )^d\ e^{-k^2r^2/4(d+1)}
\label{En2}
\end{equation}
with $\hbar k =  \sqrt{2m (E - V(\frac{{\bf x} + {\bf x'}}{2}))} $ and $r = |{\bf x} -{\bf x'}|$.  Again, the Gaussian form of this expression arises from the asymptotic limit of the Bessel function.
In the interacting case this can again be brought into the same form as the equivalent expression at constant temperature:
\begin{equation}
\rho_{cl} ({\bf x}, {\bf x'}, \beta) = \frac{1}{Z(\beta)} \left( \frac{m}{2\pi \beta \hbar^2}\right)^{d+1} e^{-\frac{m r^2}{2 \hbar^2 \beta} + V(\frac{{\bf x}+{\bf x'}}{2}) \beta}
\label{Temp2}
\end{equation}
In order to make the connection we must identify the energy with a certain temperature.  This relationship between $E$ and $\beta$ is 
\begin{equation}
E - \left< V \right> = \frac{ND}{2 \beta}
\end{equation}
where $\left< V \right> $ is the ensemble average of the potential in one of the statistical ensembles. Using this relationship in Eq.~\ref{En2} gives
\begin{equation}
\rho_{cl} ({\bf x}, {\bf x'}, E) =  \frac{1}{\rho(E)} \frac{m}{2\pi \hbar^2 d!}\left (\frac{k^2}{4\pi}\right)^d\ e^{-\frac{mr^2}{2\hbar^2 \beta}} e^{-\frac{m(\left<V\right> - V) r^2}{2 \hbar^2 (d + 1)}} 
\label{En4}
\end{equation}
In order for Eq.~\ref{En4} to be equivalent to Eq.~\ref{Temp2} the term with $\left<V\right> - V$ must be negligible.  This is true for configurations of particles which possess the typical (and vastly most probable) sum total kinetic energy for all the particles.  Since the typical total kinetic energy is by far the most probable, nearly all points in configuration space lead to small values of $\left<V\right> - V$, and that term is negligible almost always.
The remaining terms in Eq.~\ref{En4} and Eq.~\ref{Temp2} are shown to be the same by the equivalence of the classical ensembles as shown in Sec.~\ref{dl}.

It is also telling to trace over the coordinates of all but one of the interacting particles, given by a coordinate $\vec y$.  We thus seek the reduced density matrix, diagonal or off diagonal in $\vec y$.  The trace will over many coordinates be overwhelmingly dominated (in the large $N$ limit) by the most probable total kinetic energy for all the particles. 
Then we find 
\begin{equation}
G(\vec y,\vec y',\beta) \sim \lambda^{{-3N-2}} e^{-\pi r^2/\lambda^2}
\end{equation}
where $r^{2} = \vert \vec y-\vec y'\vert^{2}$ and $\lambda = h/\sqrt{2 \pi m \kappa T}$ .
Thus the quantum mechanical single particle Green function and density matrix make sense as their imaginary time counterparts in the $N \to \infty$ limit, in accordance with well known results for the canonical ensemble.



 \subsection{Large N limit and Boltzmann averaged Green functions}
Even though it is a necessary consequence of the equivalence of the ensembles, it is interesting to establish the generality of the Boltzmann average over the energy of  a noninteracting subsystem in the following way. Suppose $N-M$ particles are no longer  interacting with the remaining $M$ particles, but their states  are correlated by having been in contact in the past with the total energy fixed at  $E$. In the time domain and in an obvious notation we have 
\begin{equation} 
G_N^+({\bf y},{\bf z}; {\bf y'},{\bf z'},t)  = i \hbar \ G_{N-M}^+({\bf y},{\bf y'},t) G_M^+({\bf z},{\bf z'},t)
\end{equation} 
Then the Fourier convolution theorem can be applied to the Fourier transform into the energy domain, i.e.
\begin{equation} 
G_N^+({\bf y},{\bf z}; {\bf y'},{\bf z'},E) = \frac{i \hbar}{2 \pi} \int\limits_{-\infty}^{\infty} G_{N-M}^+({\bf y},{\bf y'}, E-E') G_M^+({\bf z},{\bf z'},E')\ dE'
\end{equation} 
which incidentally leads to some rather unlikely looking identities for Bessel functions; the reader may easily generate them.  Our purpose is served if, focussing on the subsystem of $M$ particles, we trace over the $N-M\ {\bf y}$ coordinates. This gives 
\begin{equation} 
{\rm Tr}_{\bf y}[G_{N-M}^+( E-E')]\sim  \lim_{{\bf y'} \to {\bf y}}
-\frac{m}{2 \hbar^2} \left( \frac{1}{\Gamma(d_{N-M} + 1)} \left ( \frac { {k_{N-M}}^2}{4 \pi}\right )^{d_{N-M}} + i \frac{\Gamma(d_{N-M})}{\pi^{d_{N-M} + 1} |{\bf y'}-{\bf y}|^{2d_{N-M}}} \right) 
\end{equation} 
times a volume factor, in the case of an ideal gas. The second term is not a function of $E'$.  Therefore the integral of it times $G_M({\bf z},{\bf z'},E)$ is proportional to $\delta({\bf z'}-{\bf z})$.  So long as ${\bf z} \neq {\bf z'}$ that term is zero. Neglecting all unimportant (for this argument) factors this leaves 
\begin{equation} 
{\rm Tr}_{\bf y}[G_{N-M}^+( E-E')]\propto (E-E')^{d_{N-M}} = E^{d_{N-M}}\left ( 1-\frac{E'}{E}\right )^{d_{N-M}}\sim E^{d_{N-M}}\  e^{-\beta E'}
\label{canon}
\end{equation} 
with of course $\beta = 1/\kappa T$.  In arriving at Eq.~\ref{canon} we used $E=\frac{D}{2} N \kappa T$   for the case of particles embedded in $D$ dimensions.  Finally we arrive at
\begin{equation} 
{\rm Tr}_{\bf y} [G_N^+(E)] \propto \int\limits_{-\infty}^{\infty} e^{-\beta E'} \ G_M^+({\bf z},{\bf z'},E')\ dE' =  G_M^+({\bf z},{\bf z'},\beta)
\label{bolt}
\end{equation} 
in the large $N$ limit.  This establishes the generality of the Boltzmann average over the subsystem energy for large $N$.
This discussion establishes again the connection between the canonical and microcanonical ensembles, however  in a way not involving the Bessel functions and their asymptotic form, so it is less general than other results in this paper valid  for any $N$.

\subsection{Stationary phase canonical limit}
\label{spl}
It is also possible to recover the Gaussian form in Eq.~\ref{asymp1} by carrying out the integral in Eq.~\ref{bes} by stationary phase,  provided the real factor involving $t$ in the denominator is taken into the exponent, as $-ND/2\log{t}$ i.e.
\begin{equation}
\label{spr}
G_{cl}^+({\bf x},{\bf x}+ {\bf r},E)  =  \frac{-i}{\hbar}\int\limits_0^\infty\left ( \frac{m}{2 \pi i \hbar }\right )^{ND/2} e^{i m r^2/2\hbar t - i V({\bf x} +\frac{ {\bf r}}{2})t/\hbar + i E t/\hbar - ND/2\log{t}}\ dt.
\end{equation}
The complex stationary phase point $t^{*}$  in the large $N$ limit becomes $t^{*} = -i N D \hbar /(2(E-V))$ , yielding the same result as in Eq.~\ref{asymp1}, with $ \hbar^2 k({\bf x},E)^2/2m = E-V({\bf x})$, and making this  another route between the quantum microcanonical and canonical ensembles.  
  Since the positions are arbitrary we cannot however identify the {\it average} kinetic energy with $E-V$, and thus without further averaging we cannot associate $t^*$ with any inverse temperature.  It is interesting nonetheless that there is a complex time $t^{*}$ appropriate to every position ${\bf x}$, even if that time is not related to the temperature. For an ideal gas  the stationary phase time is  $t^{*} = -i \hbar/\kappa T = -i \beta \hbar$,  after making the identification $ E = ND/2 kT$.    
A discussion about traces over most of the coordinates and the recovery of the usual  temperature through $\langle K\rangle = D/2 NkT$ proceeds as in Sec.~\ref{stl}.

\section{Constraints}

 In the large $N$ limit the ergodic hypothesis is strongly motivated,
but statistical mechanics does not pre-suppose that ergodicity is unchecked; rather  constraints are always present, such as   walls and boundaries which control volume. 
Ergodicity is then defined with respect to these constraints.
The guiding idea in this paper, i.e. the extended Berry RPW hypothesis,  is that eigenstates of the full system are  ``as random as possible, subject to prior constraints''.   In this way thermodynamic constraints arise naturally.
The real time, real energy (microcanonical ) semiclassical Green function approach not only automatically generates the averages required to get appropriate  wavefunction statistics, it also provides a natural way to include many constraints  such as walls, symmetries, and even the existence of collisions between particles  by going beyond the short time limit term to include returning (not necessarily periodic) trajectories.
The  semiclassical Ansatz for these extended problems in the presence of constraints is  
\begin{equation} 
G({\bf x},{\bf x'},t)   \approx  G_{direct}({\bf x},{\bf x'} ,t)  + \sum\limits_j G_{j}({\bf x},{\bf x'},t)  
\label{multiple}
\end{equation} 
where  $G_{j}({\bf x},{\bf x}+ {\bf r},t)  $ is a semiclassical (Van Vleck-Morette-Gutzwiller) Green function, 
\begin{eqnarray} 
\label{VanVleck}
G_{j} ({\bf x},{\bf x}^\prime;t)
 &=& \left({1\over 2\pi i\hbar}\right)^{ND/2} \bigg|{\rm Det}
\bigg( {\partial^2 S_{j}({\bf x},{\bf x}^\prime;t)\over \partial {\bf x}
\partial {\bf x}^\prime}\bigg)\bigg|^{1/2} 
 \exp 
\left(iS_{j}({\bf x},{\bf x}^\prime;t)/\hbar
-{i \pi\nu_{j} \over 2}\right) 
\end{eqnarray} 
corresponding to the $j^{th}$ trajectory contributing to the path from ${\bf x} $ to $ {\bf x}+ {\bf r} $, and $G_{direct}({\bf x},{\bf x}+ {\bf r},t) $ is given by Eq.~\ref{green}. The short time term $G_{direct}({\bf x},{\bf x}+ {\bf r},t) $, is singled out as the shortest contributing trajectory:  supposing ${\bf r}$ to be small compared to distances to walls etc., we still have a short time, ballistic trajectory as quite distinct from trajectories which have traveled some distance away and come back.  There are cases where this separation is not clean; for such cases  we can adjust notation accordingly. Note that since a trace over all position is not being taken, there is no appearance semiclassically of periodic orbits as the only surviving contributors.  ``Closed'' orbits however can  play a large role semiclassically, a fact recognized long ago by Delos\cite{delos}.

\subsection{$N$ particles and a wall}
A very useful example is provided by a plane Dirichlet  wall felt by all the particles (e.g. $\psi(\vec x_1,\vec x_2,\cdots\vec x_N) = 0$ for  $y_i =0, i=1,\cdots N)$, as in a gas confined by a  rigid container.  The Green function and eigenfunctions   must vanish if one or more particles approaches this wall. We can use the method of images, generalized to $N$ particles, if the particles are noninteracting. (The interacting case can in principle be handled by semiclassical trajectory techniques which we bring up in the next section.) 

The Green function $G_{wall}({\bf x},{\bf x'})$ will consist of the shortest distance contribution for which all particles take a direct path from ${\bf x}$ to ${\bf x'}$, plus paths where one particle has bounced off the wall, paths where two particles have, etc. These histories are included automatically if we apply the symmetrization operator which imposes the image reflections. This operator can be written 
\begin{equation} 
{\cal R} = \prod_i^N (1-R_i) = 1 -\sum\limits_i R_i + \sum\limits_{i< j} R_iR_j -\cdots
\label{product}
\end{equation} 
where $R_i$ is the operator for reflection  about the $y=0$ axis for the $i^{th} $ particle. 
Applied to the Green function $G({\bf x},{\bf x}+ {\bf r},t)$, considered as a function of the coordinates in ${\bf x}$  in the absence of the wall, ${\cal R} $ yields the series
\begin{equation} 
G_{wall}({\bf x},{\bf x'},t) = G_{direct}({\bf x},{\bf x'},t) - \sum\limits_i G_i({\bf x},{\bf x'},t) +  \sum\limits_{i<j} G_{ij}({\bf x},{\bf x'},t) -\cdots
\label{series}
\end{equation} 
where $G_i({\bf x},{\bf x'},t) $ corresponds to the $i^{th}$ particle getting from $\vec x_i$ to $\vec x_i^\prime$ by bouncing off the wall while the others take direct paths, etc. The Fourier transform gives an analogous equation for $G_{wall}({\bf x},{\bf x'},E)$.
The effect of the symmetrization is to create Green function sources reflected across the wall and given proper sign, in the manner familiar from the method of images. 
\begin{figure}
\centerline {
\includegraphics[width=6in]{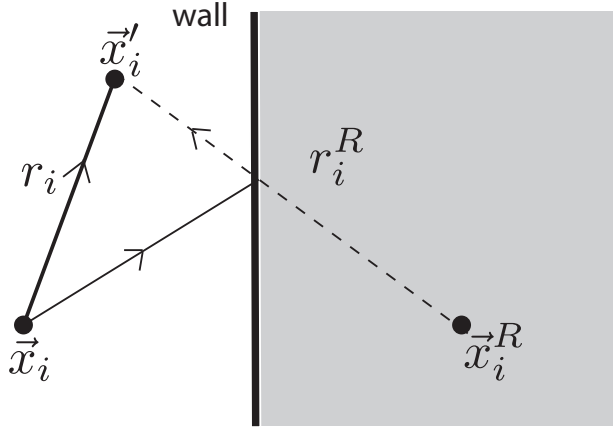}
}
\caption{A short and a bouncing path for a particle propagating near a wall. The bounce contribution, if viewed by the image method, is equivalent to a contribution of opposite sign coming from the reflected point $\vec x^R$ with the wall removed.}
\label{bounce}
\end{figure}
The short time path is shown by the direct path solid line in Fig~\ref{bounce}, corresponding to the term $G_{st}({\bf x},{\bf x'},t)  $.
The bounce path is equivalent to a  source reflected across the wall with an opposite sign, i.e. the method of images.  
Define 
\begin{equation} 
 -\frac{1}{\pi}{\rm Im} \left [  G_{st}^+({\bf x},{\bf x} +{\bf r}, E) \right ] = \frac{m}{2\pi \hbar^2}\left (\frac{k^2}{2\pi }\right )^d\ \frac{J_d(k r)}{(k r)^d} \equiv a(k) F_d( k r)
\label{main2}
   \end{equation} 
  \color{black}
Then
\begin{equation} 
 -\frac{1}{\pi}{\rm Im} \left [ G_{wall}^+({\bf x},{\bf x'},E)\right ] = a(k)\left ( F_d( k r) - \sum\limits_i F_d( k r_i)+\sum\limits_{i<j} F_d( k r_{ij}) - \cdots\right ).
 \label{allNwall}
\end{equation} 

  \color{black}
This is the general result for any $N$.   It would appear to be difficult to take it further, since all the distances, e.g. 
\begin{equation} 
r_{ij}  = \sqrt{ \sum_{m\ne i,j} \vert \vec x_m - \vec x_m^\prime\vert^2 + \vert \vec x_i^R  - \vec x_i^\prime \vert^2+\vert \vec x_j^R - \vec x_j^\prime\vert^2},
\end{equation} 
where $ \vec x_j^R$ is the reflected $j^{th}$ particle coordinates,
involve square roots.  However if we use the large $N$ asymptotic form, we find, using 
$F_d( k r)\to \exp[-k^2r^2/4(d+1)]/2^dd!$,

\begin{equation} 
 -\frac{1}{\pi}{\rm Im} \left [ G_{wall}({\bf x},{\bf x'},E)\right ] =\frac{a(k)}{2^d d!}  \prod\limits_i^N \left ( e^{- \gamma r_i^2}-e^{-\gamma  (r_i^R)^2}\right )= \frac{a(k)}{2^d d!}e^{- \gamma r^2}  \prod\limits_i^N \left ( 1-e^{-\gamma\Delta_i^2}\right )
 \label{wall}
\end{equation} 
where $\gamma = k^2/4(d+1)=\pi /\lambda^2$ and $\Delta_i^2 = (r_i^R)^2-r_i^2$.  
Since $r_i$ is the ``direct'' distance from ${\vec x_i}$ to ${\vec x'_i}$, (see Fig~\ref{bounce}),   $\Delta_i^2$ records the distance change upon reflection of the $i^{th}$ particle. We note that $\Delta_i^2$ (and thus the Green function) vanishes as any particle approaches a wall in either ${\bf x}$ or ${\bf x'}$.  It is also simple to see that the single particle density $\rho(\vec x)$ in this noninteracting case becomes, for large $N$,
\begin{equation} 
\rho(\vec x) = \rho_{0}(1-e^{-4\gamma x^{2}})
\end{equation}
where $x$ is the distance to the wall and $\rho_{0}$ is the density far from the wall.

The formulas Eq.~\ref{allNwall} and  Eq.~\ref{wall} generalize Berry's result\cite{berry2} for the wavefunction squared of one particle in two dimensions near a wall, namely 
\begin{equation} 
\langle \vert \psi(\vec x) \vert^2 \rangle =  \frac{\left(1-J_0(k | \vec x^R - \vec x | ) \right)}{\int d\vec x \left(1-J_0(k | \vec x^R - \vec x | ) \right)} .
\end{equation}
The Gaussian we get for large $N$ has a very simple interpretation.  First we note that for noninteracting systems in the canonical ensemble we can write the total density matrix as a product of one particle density matrices.  This is 
\color{black} essentially the form of Eq.~\ref{wall}, since we can write each one particle density matrix as
 \begin{equation} 
 \label{oneagain}
 \rho(\vec x,\vec x',\beta) = e^{-\gamma |\vec x - \vec x'|^2/N} \frac{ \left ( 1-e^{-\gamma(|\vec x^R - \vec x'|^2 - |\vec x - \vec x'|^2)}\right )}{ \int d\vec x \left ( 1-e^{-\gamma|\vec x^R - \vec x|^2}\right )} \to
\frac{ \left ( 1-e^{-\gamma|\vec x^R - \vec x|^2}\right )}{ \int d\vec x \left ( 1-e^{-\gamma|\vec x^R - \vec x|^2}\right )}
\end{equation} 
where the second form is the diagonal element.  
\color{black}
However Eq.~\ref{oneagain} also arises as the density matrix obtained from the Boltzmann average of Berry's result; i.e. averaging the fixed energy results over a canonical distribution of energies, as can be seen from the integral
\begin{equation} 
\frac{\int\limits_0^\infty k \ \left(1-J_0(k | \vec x^R - \vec x | ) \right) \ e^{-\beta\hbar^2 k^2/2m} dk}{\int\limits_0^\infty k \  e^{-\beta\hbar^2 k^2/2m} dk} = \left(1- e^{-m | \vec x^R - \vec x |^2/2\beta\hbar^2}  \right)
\end{equation} 
For $D=2$ and $N=1$ a Boltzmann average yields the Gaussian.  Indeed this necessarily holds in any number of dimensions; i.e. the appropriate Boltzmann average of $J_d( k r) /(k r)^d$ must yield a Gaussian for any $d$.  In the thermodynamic  $N \to\infty$ limit for noninteracting particles, each particle separately is Boltzmann distributed over energy, so the result must be the same as a Boltzmann average of the one particle results for any   dimension $D$  and for any constraints. 


\subsection{Symmetries - Fermions and Bosons}
\begin{figure}
\centerline {
\includegraphics[width=3in]{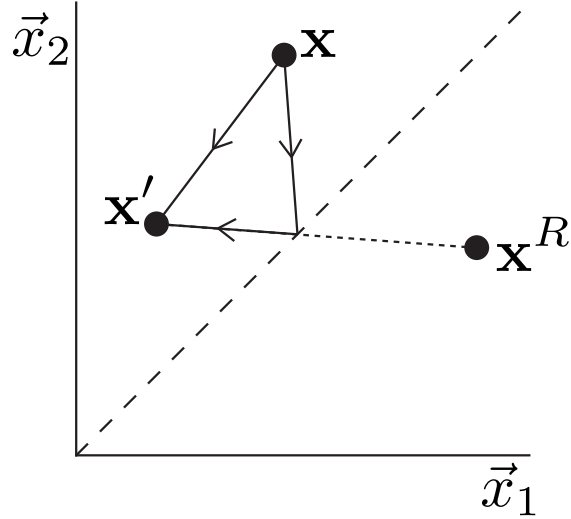}
}
\caption{The particle symmetry or antisymmetry condition is equivalent to requiring mirror symmetry or antisymmetry across the $\vec x_i = \vec x_j$ (hyper)plane.  This corresponds to having additional contributions from the images of the particles reflected over the symmetry planes.}
\label{ferm}
\end{figure}
Particle symmetry is an essential part of the many body problem. It's effect, like other symmetries, is to generate permutations where the distances have changed due to particle exchange. Figure~\ref{ferm} shows this effect graphically.  It is gratifying to see directly that permutations which induce large new distances (coming from remote pairs of particles, where ``remote'' is a relative term depending on the temperature) make little contribution. 
Consider $N$ noninteracting  Fermions or Bosons; we wish to compute the reduced density matrix for two Fermions or Bosons. This is a well known result for $N\to\infty$\cite{pathria}.   The symmetric or antisymmetric Green function is 
\begin{equation}
G_{S/A} ({\bf x},{\bf x}+{\bf r}, E) = \frac{1}{N!} \sum_n \epsilon_n \frac{-im}{2\pi \hbar^2}\left (\frac{k^2}{2\pi }\right )^d\ \frac{H_d(k r_n)}{(k r)^d} 
\end{equation}
where $r_n = \sqrt{ |\vec{x_1} - \vec{x_{p_1}}'|^2 + \cdots + |\vec{x_N} - \vec{x_{p_N}}'|^2 }$, $\{p_1, \cdots, p_N\}$ is the $n$th permutation of $\{1, \cdots, N\}$, and $\epsilon_n = 1$ if the parity of the permutation is even and $\epsilon_n = \pm 1$ if the parity of the permutation is odd (with the upper sign for bosons and the lower sign for fermions). \\
\begin{eqnarray} 
\langle \psi^\ast (\vec{x_1}\cdots\vec{x_N}) \psi(\vec{x_1}\cdots\vec{x_N}) \rangle &=& \nonumber -\frac{1}{\pi} \frac{{\rm Im} \left( G_{S/A} ({\bf x},{\bf x}+{\bf r}, E) \right)}{\rho(E)}
   \\ &=& \frac{1}{\rho(E)N!} \sum_n \epsilon_n \frac{m}{2\pi \hbar^2}\left (\frac{k^2}{2\pi }\right )^d\ \frac{J_d(k r_n)}{(k r)^d} 
\end{eqnarray} 
In the limit that $N$ is large, this becomes
\begin{equation}
\langle \psi^\ast (\vec{x_1}\cdots\vec{x_N}) \psi(\vec{x_1}\cdots\vec{x_N}) \rangle =  \frac{1}{\rho(E)N!} \sum_n^{N!} \epsilon_n   \frac{m}{2\pi \hbar^2 d!}\left(\frac{k^2}{4 \pi}\right )^d\ e^{-k^2r_n^2/4(d+1)}
\end{equation}
The diagonal component of this with the $r_n$'s written out explicitly is
\begin{equation}
\langle \psi^\ast (\vec{x_1}\cdots\vec{x_N}) \psi(\vec{x_1}\cdots\vec{x_N}) \rangle =  \frac{m}{2\rho(E)N! \pi \hbar^2 d!}\left(\frac{k^2}{4 \pi}\right )^d \sum_n^{N!} \epsilon_n   \ e^{-k^2(\vec x_1 - \vec x_{p1}) ^2/4(d+1)} \cdots e^{-k^2(\vec x_N - \vec x_{pN}) ^2/4(d+1)}
\end{equation}
Up to the normalization constant this is the constant temperature density matrix for N noninteracting fermions or bosons:
\begin{equation}
\langle \psi^\ast (\vec{x_1}\cdots\vec{x_N}) \psi(\vec{x_1}\cdots\vec{x_N}) \rangle  =  \frac{m}{2\rho(E)N! \pi \hbar^2 d!}\left(\frac{k^2}{4 \pi}\right )^d \sum_n^{N!}  \epsilon_n \ e^{-m(\vec x_1 - \vec x_{p1})^2/2\beta \hbar^2} \cdots  e^{-m(\vec x_N - \vec x_{pN})^2/2\beta \hbar^2}
\end{equation}
Again the identification $E = \frac{D}{2}N \kappa T$ was used.
This can be rewritten as an integral over wavevectors:
\begin{equation}
\langle |\psi({\bf x})|^2 \rangle = A  \sum_n^{N!} \epsilon_n \int d\vec k_1 \cdots d\vec k_N \ e^{-\beta \hbar^2 {k_1}^{2}/2m + i\vec k_1 \cdot(\vec x_1 - \vec x_{p1})} \cdots e^{-\beta \hbar^2 {k_N}^{2}/2m + i\vec k_N \cdot(\vec x_N - \vec x_{pN})}
\end{equation}
where $A = \frac{m}{2\rho(E)N! \pi \hbar^2 d!}\left(\frac{k^2}{4 \pi}\right )^d \left(\frac{\beta \hbar^2}{2 \pi m}\right)^{d+1}$ is the normalization constant.
 Rearranging gives
\begin{equation}
\langle |\psi({\bf x})|^2 \rangle = A \sum_n^{N!}  \epsilon_n \int d\vec k_1 \cdots d\vec k_N \ e^{-\beta \hbar^2 ({k_1}^{2} + \cdots + {k_N}^{2})/m} e^{i(\vec k_1 - \vec k_{p1}) \cdot \vec x_1} \cdots e^{i(\vec k_N - \vec k_{pN}) \cdot \vec x_N }
\end{equation}
If the volume that the particles are confined to is large but finite,
\begin{equation}
\int \langle |\psi({\bf x})|^2 \rangle  d\vec x_3 ... d\vec x_N = A V^{N-2}  \sum_n^{N!}  \epsilon_n  \int d\vec{\bf k} \ e^{-\beta \hbar^2 {\bf k}^2/2m} e^{i(\vec k_1 - \vec k_{p1}) \cdot \vec x_1 } e^{i(\vec k_2 - \vec k_{p2}) \cdot \vec x_2 } \delta_{\vec k_3, \vec k_{p3}} \cdots \delta_{\vec k_N, \vec k_{pN}}
\end{equation}
For fermions if the wavevector of any two particles are the same the term is killed by the term with the wavevectors reversed in accordance with the Pauli principle.  This leaves only two terms
\begin{equation}
\int \langle |\psi({\bf x})|^2 \rangle d\vec x_3 \cdots d\vec x_N = A V^{N-2} \sum_n^{N!} \epsilon_n   \int d{\bf k} \ e^{-\beta \hbar^2 {\bf k}^2/2m} e^{i(\vec k_1 - \vec k_{p1}) \cdot \vec x_1} e^{i(\vec k_2 - \vec k_{p2})\cdot \vec x_2 }   
\end{equation}
For bosons there are also only two types of terms, but each is multiplied by the same factor since like terms are added together.
Either way, carrying out the integral over ${\bf k}$,
\begin{equation}
\int \langle |\psi({\bf x})|^2 \rangle d\vec x_3 \cdots d\vec x_N = \frac{\left( 1 \pm e^{-m(\vec{x_1} - \vec{x_2})^2/ \beta \hbar^2}\right)}{\int d\vec x_1 d\vec x_2 \left(  1 \pm e^{-m(\vec{x_1} - \vec{x_2})^2/ \beta \hbar^2}\right)}  
\end{equation}
This is the well known result for the density of two noninteracting fermions or bosons.


\section{Scattering}
\label{pot}
A hard wall is a potential energy feature which induces a boundary condition, requiring the wavefunction or Green function to vanish as the wall is approached. Softer potentials do not induce fixed boundary conditions and require a different treatment.  A potential may still however be thought of as a constraint: we consider waves as random as possible subject to the existence of a potential, be it fixed or interparticle. In practice this means we return to the Green function formulation used throughout.

Consider a  soft repulsive or attractive potential somewhere in a noninteracting gas.  
Assuming no boundaries, mutually noninteracting particles can interact with the potential 0 or 1 times. (We assume for simplicity that the potential is short ranged.  Because of the ergodicity assumption inherent to the random wave hypothesis, the presence of remote walls would actually make no difference.) This circumstance develops along lines very similar to the wall, except that we cannot use the method of images.  It illustrates the use of the full semiclassical propagator within this formalism.

\begin{figure}
\centerline {
\includegraphics[width=6in]{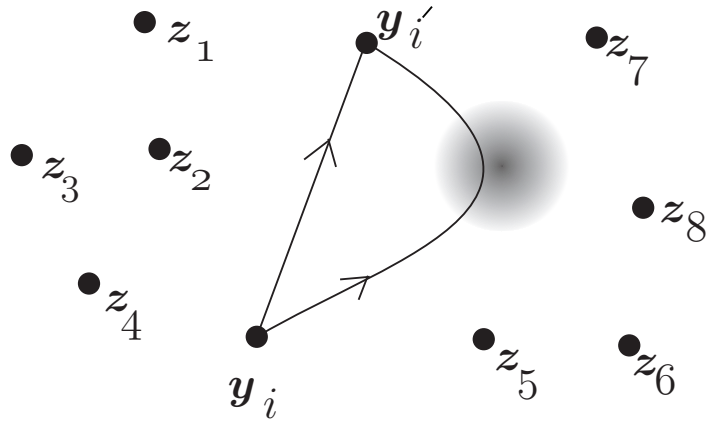}
}
\caption{A short ballistic and a colliding path both lead to the same final point for a particle propagating near a localized repulsive potential. The colliding path cannot be treated by the short time approximation; rather, a Van Vleck Green function is required. In this term, all but the $i^{th}$ particle remain in place.}
\label{bounceP}
\end{figure}

Eq.~\ref{product} and Eq.~\ref{series} both hold, with the effect of $R_i$ changed to mean ``the $i^{th}$ particle takes the path from initial to final coordinates in  which it  deflects from the potential, if such a path exists classically''.  For $N$ particles, there is a ``direct'' term in Eq.~\ref{series} where no particle interacts with the potential, $N$ terms where one of them does, etc. 
We have, in the simple case shown in Fig.~\ref{bounceP}, and in analogy with Eq.~\ref{series},
\begin{equation} 
G({\bf x},{\bf x'},t) = G_{direct}({\bf x},{\bf x'},t) + \sum\limits_{i}G_{bounce,i }({\bf x},{\bf x'},t) +\sum\limits_{i,j}G_{bounce,i,j }({\bf x},{\bf x'},t) +\cdots
\label{series2}
\end{equation} 
with $G_{direct}({\bf x},{\bf x'},t) $ given by Eq.~\ref{green}, and e.g. 
\begin{equation}
G_{bounce,i}({\bf z,y}_{i},{\bf z}+ {\bf r},{\bf y}_{i}^{\prime},t)  
\approx \left ( \frac{m}{t}\right )^{\frac{(N-1)D}{2}} \left(\frac{1}{2 \pi i \hbar }\right)^{\frac{ND}{2}}\left| \partial^2 S_i({\bf y}_{i},{\bf y}_{i}^\prime;t)\over \partial {{\bf y}_{i}\partial {\bf y}_{i}^\prime}\right |^{\frac{1}{2}}  e^{i m r^2/2\hbar t - i V({\bf z} +\frac{ {\bf r}}{2})t/\hbar +
 iS_{i}({\bf y}_{i},{\bf y}_{i}^\prime;t)/\hbar
-{i \pi\nu_{i} \over 2} }
\label{tog}
\end{equation}
Considering this
term where only the $i^{{th}}$ particle with coordinate ${\bf y}_{i}$ interacts with the potential,   we have $N-1$ ``spectator''  ${\bf z}$ particles, and the propagator becomes a product of the noninteracting Green function for $N-1$ particles and a more complicated Van Vleck semiclassical term for the colliding particle. 
The noninteracting part contributes a term  $(N-1)D/2\log{t}$  in the exponent along with the one particle classical action of the $i^{th}$ particle. For sufficiently large $N$, and tracing over the ${\bf z}$ particles, this factor leads again to the usual time condition $t^{*} = -i \beta \hbar$ and a thermal average of the one particle energy Green function under the Fourier transform from time to energy, as in  Equation~\ref{bolt}:
\begin{equation} 
G({\bf y},{\bf y'},E) \approx G({\bf y},{\bf y'},\beta )= G_{direct}({\bf y},{\bf y'},\beta) + \sum\limits_{i}G_{bounce,i }({\bf y},{\bf y'},\beta) +\sum\limits_{i,j}G_{bounce,i,j }({\bf y},{\bf y'},\beta) +\cdots
\label{seriesbeta}
\end{equation} 

$t^{*} = -i \beta \hbar$  becomes the imaginary time over which the action for the ${\bf y}$ coordinates are evaluated.




\section{Conclusion}

Starting with Berry's random plane wave conjecture for chaotic Hamiltonian systems, we have followed it's implications for moderate and large numbers of particles $N$. In the large $N$ limit we have necessarily  arrived at some familiar territory in statistical mechanics. We have adopted  a Green function, semiclassical perspective, arriving at a Gaussian-Bessel function asymptotic result for energy Green functions, providing an {\it analytic}  connection between the quantum microcanonical and canonical ensembles.  We have extended the incorporation of constraints into the random wave hypothesis, considering several types of constraints, including walls and interparticle collisions.   
Indeed the guiding perspective has been to make quantum waves ``as random as possible subject to known prior constraints''.   This must ultimately be  equivalent to the ergodic hypothesis of quantum statistical mechanics.  The nonstandard  methods and perspective used here may possibly lead to new avenues of inquiry, and it is our hope that the semiclassical approach might permit new ways of treating strongly interacting systems.
 
 The next stage in the development of this approach is to consider short ranged potentials between particles, i.e. interparticle collisions.  The first corrections to 
 the free particle limit involve binary collisions, which  can be computed semiclassically or using a delta potential appropriate to s-wave scatterers.  Again the effect of the other particles will be to provide a thermal reservoir which essentially averages the Green function over a thermal distribution of energies (if N is sufficiently large).  We save this for a future paper, where we hope to examine specific potentials and derive two particle radial distribution functions.




{ Acknowledgments }
We thank Adam Wasserman for helpful discussions, and the National Science Foundation under grant NSF-CHE- 
0073544.

 \end{document}